\documentclass[twocolumn,prl,showpacs,amsmath,amssymb,floatfix,superscriptaddress]{revtex4-1}
\usepackage[latin9]{inputenc}
\usepackage{color}
\usepackage{graphicx}
\usepackage{esint}

\makeatletter

\providecommand{\tabularnewline}{\\}

\definecolor{zgreen}{rgb}{0.0, 0.196, 0.15}\pdfoutput=1

\usepackage{dcolumn}
\usepackage{bm}

\makeatother

\begin{document}
\title{Exploring Metamagnetism in Triangular Ising Networks: Insights from
Further-Neighbor Interactions with a Case Study on ErGa2}
\author{Po-Hao Chang}
\email{pchang8@gmu.edu}
\author{Igor I. Mazin}
\email{imazin2@gmu.edu}
\affiliation{Department of Physics and Astronomy, George Mason University, Fairfax,
VA 22030, USA}
\affiliation{Quantum Science and Engineering Center, George Mason University, Fairfax,
VA 22030, USA}
\begin{abstract}
The classical Ising model on the triangular lattice (we will call
it I-3 model below), while simple in the nearest-neighbors (NN) only
approximation, becomes increasingly richer and more complex when further
interactions are incorporated. However, the studies so far have not
been exhaustive, nor have any attempts been made to estimate how realistic
are the parameter ranges that generate strong metamagnetism with a
large number of magnetization steps. In this study, we identify one
such candidate, ErGa$_{2}$, a material known to have one strong magnetization
step, albeit some narrow steps below and above cannot be confidently
excluded. It has been established, and we can confirm the same computationally,
to have an easy axis perpendicular to the triangular Er plane, with
a strong anisotropy and with a large magnetic moment of $9.5\:\mu_{B}$,
making it a perfect implementation of the classical I-3 model. In
the first part of the analysis, we present the I-3 model with up to
the third nearest-neighbors in a range of parameters $J_{2}$ and
$J_{3}$ ($J_{1}$ in this part is set to 1), and in some cases adding
a rather small $J_{4}$ in order to reveal new phases otherwise degenerate
with some others. The richest phase diagram is, not surprisingly, observed
when all interactions are antiferromagnetic (AF). Subsequently, a
more realistic case, inspired by RKKY and by our calculations for
ErGa$_{2}$, where $J_{1},J_{2}>0$ (antiferromagnetic) and $J_{3},J_{4}<0$
(ferromagnetic), is presented. Finally, we report our first-principles
calculations of $J_{1-4}$ in ErGa$_{2}$ and compared the phase diagram
in the regime corresponding to the calculated values with the experiment.
\end{abstract}
\maketitle

\section{Introduction}

There has been recent interest in 2-dimensional magnetic systems 
as they are often candidates for nontrivial magnetic states, such
as spin liquids \cite{Balents2010,Santos2004,Han2012,Nakatsuji2005}.
Considerable progress has been achieved regarding insulating 2D
Kagome lattices dominated by short-range interactions \cite{Norman}. Magnetic
anisotropy in such systems is small, and the continuous degeneracy
of magnetic order plays an important role.

A relatively recent addition to this landscape are rare-earth-based
systems \cite{hoagge2020} with a strong uniaxial anisotropy, dominating
over a relatively weak exchange. Interesting physics there comes from
the fact that, on one hand, the Ising model on a Kagome lattice does
not show a phase transition (as opposed to the famous square lattice
\cite{landau2013statistical}), and, on the other hand,
when embedded in a good metal background the rare earth ions show
a relatively weak, but long-range and potentially sign-changing exchange
interaction (related, of course, to Rudermann-Kittel-Kasuya-Yosida,
RKKY, interaction \cite{Kittel1954,Kasuya1956,VanVleck1962}). This
leads to these materials exhibiting a number of metamagnetic transitions,
that is, discrete steps in magnetization as a function of magnetic
field, ranging from two to as many as nine steps.

In this case, the strong geometric frustration associated with Kagome
lattice is not necessary (albeit may be helpful). The simple triangular,
not really frustrated in Heisenberg or XY models, where the ground
state is uniquely defined as a 120$^{\circ}$ spin star, is frustrated
in the sense of an infinitely degenerate discontinuous ground state, in the classical
Ising model \cite{Wannier1950}.

The classical Ising model on the triangular lattice
while simple in the nearest-neighbors (NN) only approximation,
becomes increasingly richer and more complex when further interactions
are added \cite{Tanaka1975,METCALF1974,Chen2016,Uryu1975}. For instance,
in Ref. \cite{Tanaka1975,METCALF1974,Chen2016,Uryu1975} possible
ground state orders were identified, upon including up to the fifth
nearest-neighbors. However, the studies so far have not been exhaustive,
nor any attempt was made to estimate how realistic the parameter
ranges that generate strong metamagnetism with a large number of magnetization
step.

In this paper, we identify one such candidate, ErGa$_{2}$, a material
known since 1970s \cite{Tsai1978,Tsai1979} and having one strong magnetization
step \cite{Doukourb1982,Tsai1978,Tsai1979,dosreis2014,dosreis2010},
albeit some narrow steps below and above cannot be confidently excluded.
It is known, and we can confirm the same computationally, to have
an easy axis perpendicular to the triangular Er plane, with a strong
anisotropy \cite{DIVIS1997L81,Ball1995}, with the large magnetic moment of $9.5\:\mu_{B}$ \cite{Doukourb1982,Tsai1978,Tsai1979,dosreis2014,dosreis2010},
making it a perfect implementation of the classical I-3 model.

The paper is organized as follows. First, since we believe that,
despite a number of papers on long-range triangular Ising the full
phase diagram has not been established and its complexity is not appreciated,
in the first part, we study the I-3 model with up to the 3rd nearest
neighbors in a range of parameters $J_{2}$ and $J_{3}$ ($J_{1}$
in this part is set to 1), and in some cases adding a rather small
$J_{4}$ in order to reveal a new phase otherwise degenerate with some
others.  The richest phase diagram is observed when all interactions
are antiferromagnetic (AF); this case is discussed in Section \ref{sec:Geneneral}.
Subsequently, a more realistic case, inspired by RKKY and by our calculations
for ErGa$_{2}$, where $J_{1},J_{2}>0$ (antiferromagnetic) and $J_{1},J_{2}<0$
(ferromagnetic), is presented. Next, we present our first-principles
calculations of $J_{1-4}$ in ErGa$_{2}$ and compared the phase diagram
in the regime corresponding to the calculated values (as well as ``around''
them, to account for possible inaccuracy in DFT calculations).

\section{General Model\label{sec:method}}

\begin{figure*}
\centerline{\includegraphics[width=0.9\linewidth]{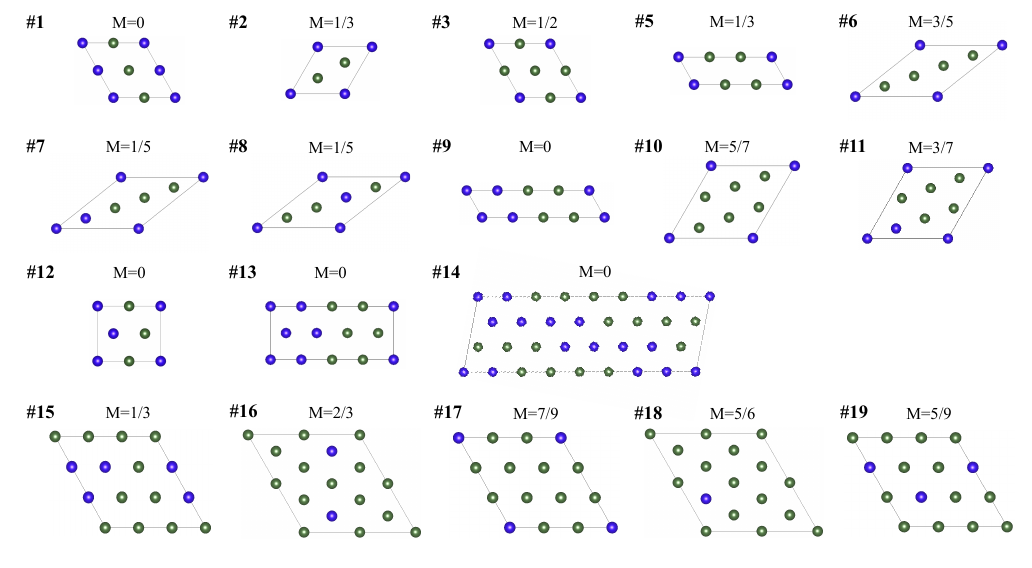}}

\caption{\label{fig:configs}All the possible magnetic ground states with their corresponding magnetization per spin. Blue and green colors indicate 
opposite spins. The ferromagnetic state labeled as \#4 is not listed.}
\end{figure*}

The model magnetic Hamiltonian of up to fourth-nearest-neighbor (4NN)
in the presence of an external magnetic field is as follows: 
\begin{equation} 
\begin{split} 
H &=\sum_{\left\langle ij\right\rangle _{1}}J_{1}S_{i}^{\mathit{z}}S_{j}^{\mathit{z}}+\sum_{\left\langle ij\right\rangle _{2}}J_{2}S_{i}^{\mathit{z}}S_{j}^{\mathit{z}}\\   &+ \sum_{\left\langle ij\right\rangle _{3}}J_{3}S_{i}^{\mathit{z}}S_{j}^{\mathit{z}} +\sum_{\left\langle ij\right\rangle _{4}}J_{4}S_{i}^{\mathit{z}}S_{j}^{\mathit{z}}-h\sum_{i}S_{i},
\label{eq:spin_ham} 
\end{split} \end{equation}
where $S$ are the normalized Er moments, ($\mathbf{S}=\mathbf{M}(Er)/M(Er)$, 
$|\mathbf{S}|=1$, and $h=H_{ext}M(Er)$ is Zeeman energy in the external field $H_{ext}||z$.
Since finding a new ground state is generally not a trivial task, particularly
when more further-neighbors are included, in our study, 
we implemented a similar approach to Ref. \cite{HORIGUCHI19729}.
We summarized the possible ground state orderings that have been suggested in the literature \cite{Chen2016,Tanaka1975,Uryu1975,METCALF1974}
and then the phase diagrams were obtained by identifying the configuration
with the lowest energy for every parameter set (i.e., $h$ and $J_{i}$).
All the magnetic states discussed in this study are labeled with
a number preceded by the ``\#'' sign as defined in Fig. \ref{fig:configs}.

In our first-principles-based analysis, we calculated the total energies
for six different magnetic orderings, \#1, \#2, \#3, \#5, \#9 and
\#12 as shown in Fig. \ref{fig:configs}. The data were then fitted
to Eq. \ref{eq:spin_ham} to extract exchange parameters $J_{1-4}^{\mathrm{DFT}}$. 

The total energy for each configuration was calculated using Vienna
ab initio Simulation Package (VASP) \cite{Kresse1996} within the projector
augmented wave (PAW) method \cite{blochl1994,Kresse1999}. The Perdew-Burke-Enzerhof
(PBE) \cite{PBE} generalized gradient approximation was employed
to describe exchange-correlation effects. 

The on-site Coulomb interactions are taken into account using LDA+U
\cite{Anisimov1991} to improve the description of the interactions
between the localized $f$-electrons of Er. A large $U-J=8$ is used.
The experimental lattice parameters $a=4.1861$ and $c=4.0187$ \(\text{\AA}\)  taken
from Ref. \cite{Tsai1979} are used and are fixed for all the calculations.

\section{Geneneral Discussion\label{sec:Geneneral}}

\subsection*{Case 1: $J_{1}$, $J_{2}$, $J_{3}$ and $J_{4}>0$ }

In the first part, we consider a more frustrated case where all the
magnetic moments are antiferromagnetically couple to each other (i.e
all $J_{i}>0$) with the 4NN only added to lift the observed degeneracy.
A detailed analysis in similar setups with more further-neighbors has been done
in the previous work \cite{Chen2016}. 

\begin{figure}
\centerline{\includegraphics[scale=0.32]{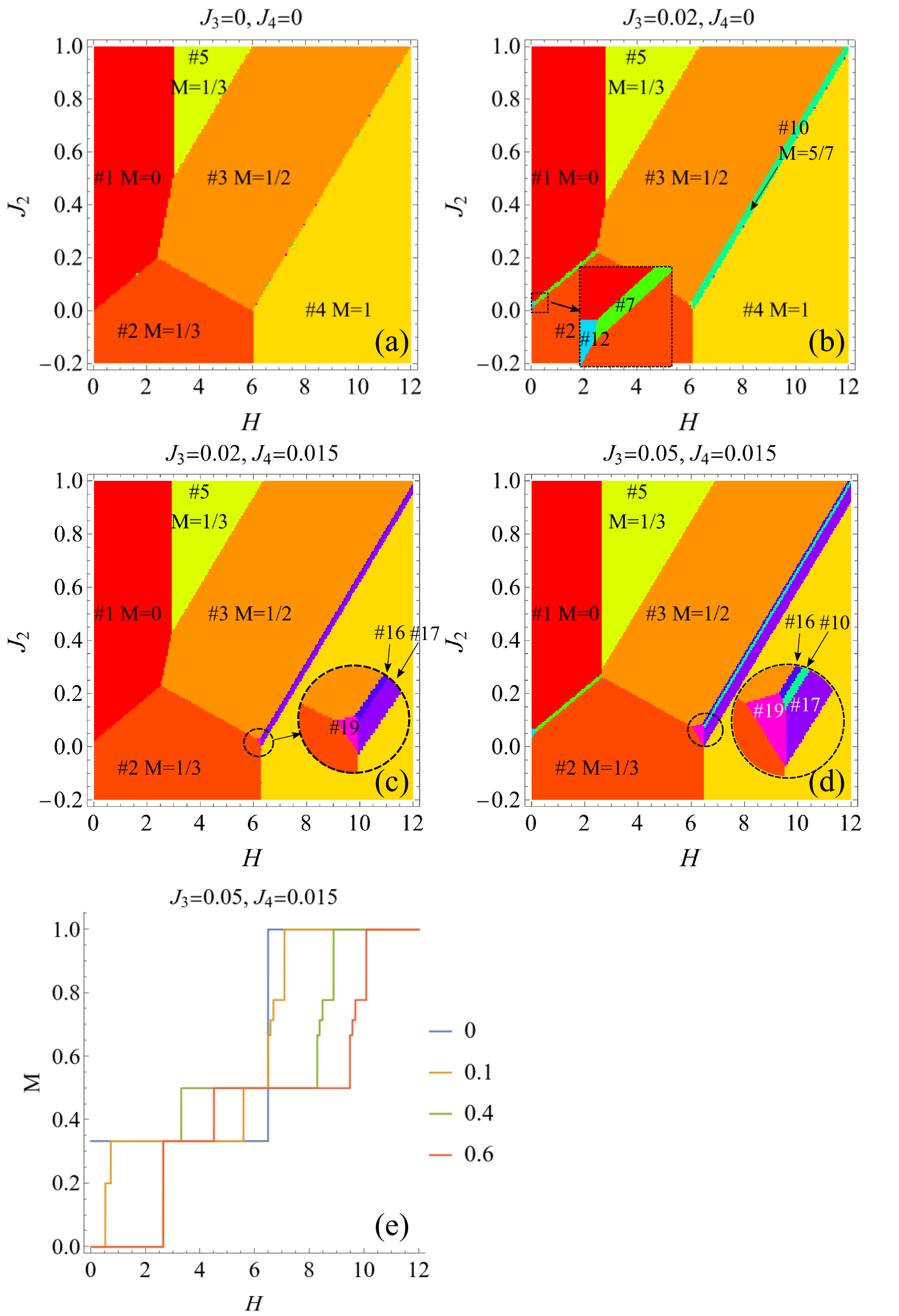}}

\caption{\label{fig:j_af}(a)-(d) show the phase diagrams for different weak
$J_{3}$ and $J_{4}$ and (e) M-H curve for $(J_{3},J_{4})=(0.05,0.015)$
at several different $J_{2}$ as specified in the legend.}
\end{figure}
Figure. \ref{fig:j_af}(a) shows the phase diagram for $J_{3}=0$
which corresponds to the 2NN case. There are five major stable phases
in the given parameter space. Two exist without an external field, $h$, and
three are induced by $h$. The  ground state at the low field is stripe
AF and borders with \#2, \#3 or \#5, depending on the strength of
$J_{2}$. Along the boundaries separating \#1 and \#2, as well as
\#3 and \#4, there are also subtle traces of multiple degenerate states,
where a few very tiny points corresponding to phases such as \#10 or \#17
can be discerned. 

To investigate the behavior of the possible multifold degeneracies,
we first include a small finite $J_{3}=0.02$ as shown in Fig. \ref{fig:j_af}(b).
As a result, two very fine straight lines emerge. One corresponds
to \#7 which lies between \#1 and \#2 phases suggesting a possible triple
degeneracy. It is worth noting that, at zero field, there is also
a different type of AF ground state \#12. The other line (\#10) lies
between \#3 and \#4 suggesting a possible quadruple degeneracy. Although
the presence of $J_{3}$ partially removes the degeneracies between
\#3 and \#4, the red dots that correspond to \#17 remain degenerate
at the \#4 and \#10 border. This state corresponds to the total magnetization
$M=7/9$ and has been observed and discussed in Ref. \cite{Chen2016}
as its formation requires introducing $J_{4}$. 

Subsequently, if another small next further-neighbor $J_{4}=0.015$ is
introduced, the state \#7 and \#10 disappear (see Fig. \ref{fig:j_af}(c)).
The disappearance of the former is due to the expansion of the \#2 phase
and the latter is replaced with new phases \#16 and \#17. When $J_{3}$
increases further to $0.05$, both states (\#7 and \#10) reappear
and a small region of new phase \#19 can also be seen in Fig. \ref{fig:j_af}(d).
Interestingly, the long narrow belt-shaped area between \#3 and \#4,
consists of three parallel thin lines across nearly the entire range of $J_{2}$.
These thin lines, resulting from the lifting of degeneracy by weak
interactions with further-neighbors, appear in the the M-H plots as
three successive short steps illustrated in Fig. \ref{fig:j_af}(e),
which depicts the field-dependent magnetization for $J_{2}=0$ and
three different $J_{2}$ values.  The shape of these short steps, resulting from the
further-neighbor interactions, does not vary with $J_{2}$ except
for $J_{2}=0$ and persists throughout nearly the entire range of
$J_{2}\neq0$. 

On the other hand, there is also another short step in the small
field region as indicated by the yellow line ($J_{2}=0.1$) as a consequence
of $J_{3}$ lifting the degeneracy between \#1 and \#2. These trends
suggest some delicate competition between $J_{3}$ and $J_{4}$ as
each might favor particular orderings and many of the states are very
close in energy.\textbf{ }These short steps are likely to exist
in a system with weak further-neighbors, but could easily get washed
out in an experiment due to defects in the sample.\textbf{ }In the
later discussion, we will only consider a small fixed $J_{4}=0.015$
simply to lift the obvious degeneracy.

\begin{figure}
\centerline{\includegraphics[scale=0.32]{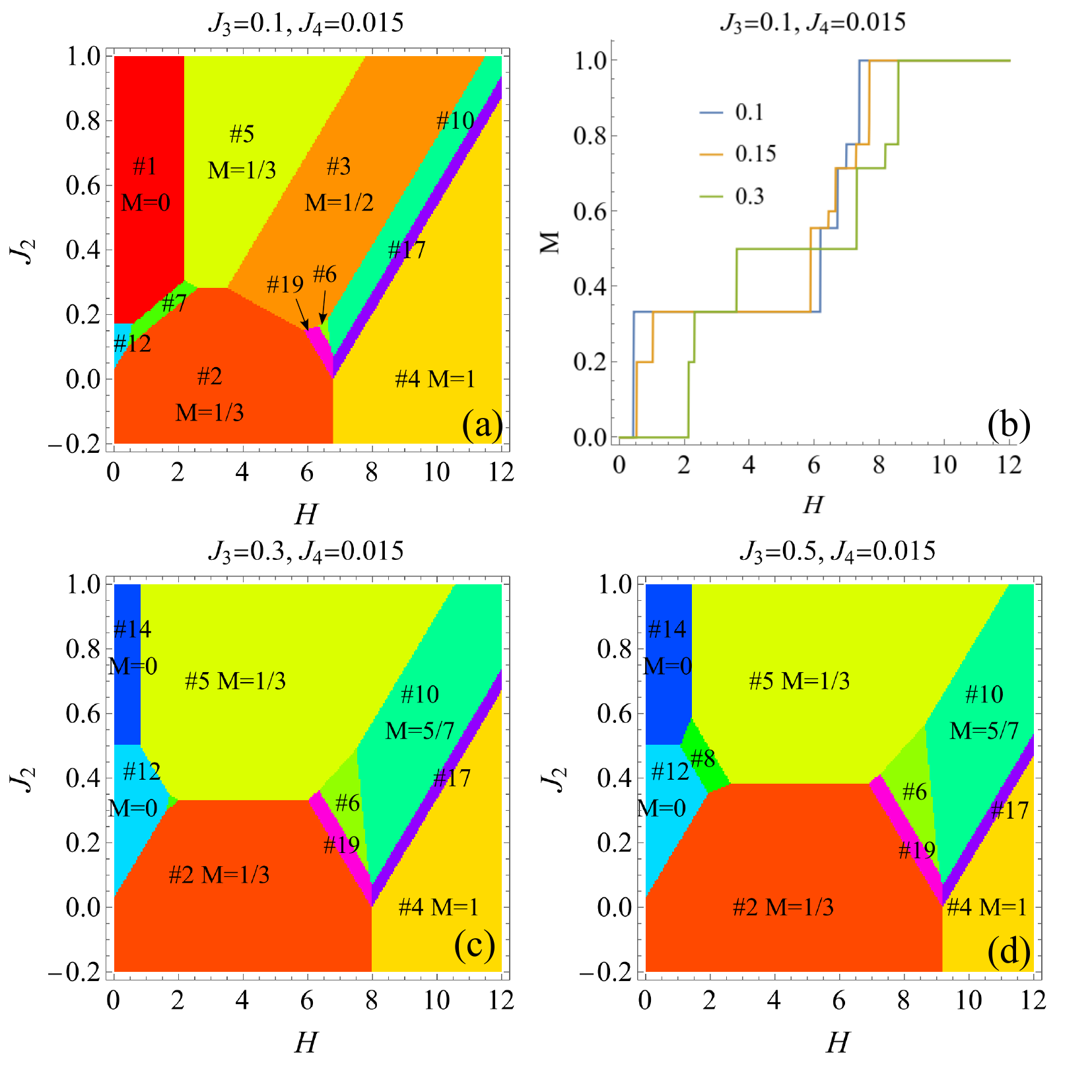}}

\caption{\label{fig:j_af2}(a), (c) and (d) show the phase diagrams for different
weak $J_{3}$ and $J_{4}$ and (b) M-H curve for $(J_{3},J_{4})=(0.05,0.015)$
at several different values of $J_{2}$.}
\end{figure}

As $J_{3}$ increases to $0.1$ as shown in Fig. \ref{fig:j_af2}(a)
the phases \#7, 10, 12, and 19 continue to expand, and new phases \#6
emerges. In the low field region, the range of stability of \#1 shrinks.
For larger $J_{2}$ ($>0.3$), it requires less field to induce the
transition to \#5 state and for the region of small $J_{2}$, it is
directly replaced by \#12 (an alternative AF) and \#7. On the other
hand, along the 3-4 boundary, the
long stripe area (\#10) and the lime green triangular region (\#6)
strongly favored by $J_{3}$ are stabilized and expanding, with increasing
$J_{3}$, at the cost of \#3 and \#4. The trace of \#17 state which
still persists at the \#4 boundary does not seem to be affected by
a moderate $J_{3}$. Fig. \ref{fig:j_af2}(b) shows the M-H curves
for there different values of $J_{2}$ with $J_{3}/J_{4}=0.1/0.015$.
The field dependency is rather sensitive to $J_{2}$. While
all three curves have rather complicated transition steps, the most
rich transition behavior happens when the magnitude of $J_{2}$ is
roughly comparable to $J_{3}=0.1$. As discussed earlier, this is
the result of the competition between different neighbors which in
turn leads to a rich phase diagram in the M-H curve that can contain
as many as seven transition steps.
The phase diagrams for $J_{3}=0.3$ and $0.5$ are shown in Fig.\ref{fig:j_af2}(c)
and (d) respectively. A few trends as the consequences of these large
$J_{3}$ can be summarized as follows. Begin with \#1, an AF state that
is disfavored by $J_{3}$ has been completely replaced by a less common
new AF phase \#13 for larger $J_{2}$ and the phase expands as $J_{3}$
increases. This phase has been discussed analytically in the early
study by Tanaka \cite{Tanaka1975}. A similar pattern, the shrinkage
of \#7, can also be found that is eventually replaced by a new phase
\#8 as $J_{3}$ continue to increase.

\subsection*{Case 2: $J_{1}$, $J_{2}>0$ and $J_{3}$, $J_{4}<0$ }

Inspired by our DFT calculations for the ErGa$_{2}$ systems, reported in the next Section, this
second scenario is considered to mimic longer-range RKKY-type exchange
couplings where the sign oscillates with distance. Interestingly,
unlike the more frustrated first scenario, the third NN $J_{3}<0$
alone does not introduce any new phase within the parameter range
of interest. We then consider up to the fourth NN where both $J_{3}$
and $J_{4}<0$ (FM). This setup is also more consistent with our DFT
data, where the model provides an excellent fitting quality. 

\begin{figure*}
\centerline{\includegraphics[scale=1.25]{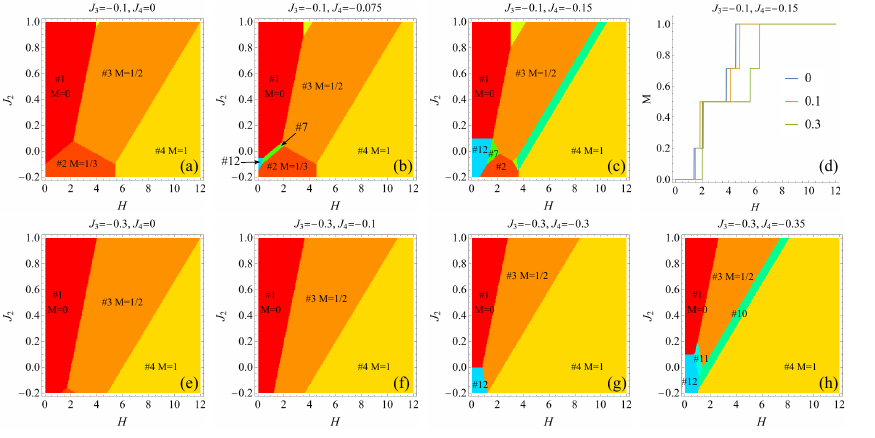}}

\caption{\label{fig:j_rkky1}Phase diagrams and magnetization vs applied field.}
\end{figure*}

In this case, we attempt to explore the behavior in the parameter
space around the exchange coupling parameters that are extracted from
our first-principles calculations. The phase diagrams are summarized
in Fig. \ref{fig:j_rkky1}, where the top row (a)-(c) and bottom row
(e)-(h) correspond to $J_{3}=-0.1$ and $-0.3$ respectively and for
each $J_{3}$, several selected $J_{4}$ values are considered.

We first isolate the effect of $J_{3}$ by comparing Figs. \ref{fig:j_rkky1}(a),\textbf{
}(e) and Fig. \ref{fig:j_af}(a). We found that $J_{3}$ alone does
not introduce any new phases but shifts the phase boundaries in favor,
strongly, of the phases \#1, \#3, and \#4. Due to the ferromagnetic
nature of $J_{3}$, which reduces the frustration in the system,
the multiple degeneracies along the \#3 and \#4 boundary no longer
persist. 
In Fig. \ref{fig:j_rkky1}, both top and bottom rows (i.e. $J_{3}=-0.1$
and $-0.3$) exhibit similar trends, in the sense that with increasing
$J_{4}$, new phases are developing in the region characterized by
small $J_{2}$ and low magnetic field, as well as along the \#3 and
\#4 boundary.\textbf{ }This again shows some competition between $J_{3}$
and $J_{4}$ and one would expect rich phase diagrams  when $J_{4}$
and $J_{3}$ are comparable.

The configurations in the top row (i.e. Fig. \ref{fig:j_rkky1}(a)-(c))
are particularly of interest since a small $J_{3}$ is more likely
to happen in a real system. When $J_{4}=-0.075$ is more or less comparable
to $J_{3}$ (see Fig. \ref{fig:j_rkky1}(b)), new phases (\#7 and
\#12) begin to emerge between \#1 and \#2, similar
to those shown in Fig. \ref{fig:j_af}(b) but with the boundaries slightly
pushed down due to the effect of a positive $J_{3}$. This thin stripe
\#7, again, indicates the possibility of having an additional short
step in the M-H curve for small $J_{2}$ (see Fig. \ref{fig:j_rkky1}(d)).
On the other hand, with a slightly larger $J_{4}=-0.15$, both \#7
and \#12 expand and a new phase appears along \#3--\#4 border.
In this case, one can see as many as four transitions (five phases) in the
M-H curve as shown in Fig. \ref{fig:j_rkky1}(d).
It is worth noting that, while all three curves begin with zero magnetization
in the small fields, the ordering for $J_{2}>0.1$ is in fact zigzag type AF phase (\#12).

In the bottom row (i.e. Fig. \ref{fig:j_rkky1}(e)-(h)), a larger $J_{3}$
further stabilizes \#1, \#3 and \#4, and in this case, at the cost
of \#2 as shown in Fig. \ref{fig:j_rkky1}(e) and it would require
a larger $J_{4}$ to introduce new phases as shown in Fig. \ref{fig:j_rkky1}(f)
and (g). For $J_{4}=-0.35$, the zigzag phase which strongly favored
by $J_{4}$ expands into positive $J_{2}$ region. Interestingly,
another new phase $M=3/7$ also develops, a phase that does not exist
in the previous setting where all $J_{i}>0$.

\section{A Case Study: ErGa2}

\begin{figure}
\centerline{\includegraphics[scale=0.29]{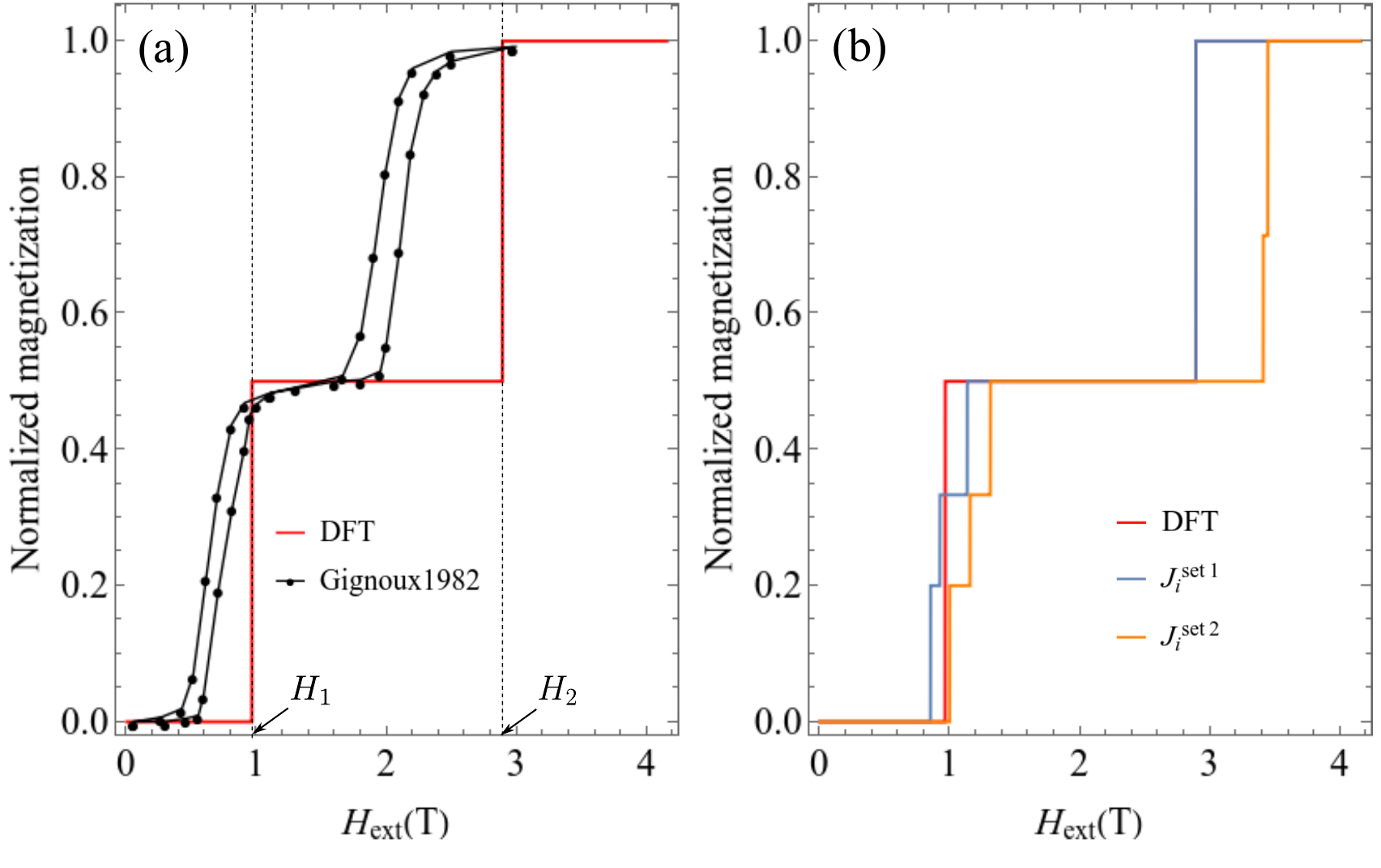}}

\caption{\label{fig:erga2}The comparison of field-dependent magnetization
between (a) DFT and experimental data taken from Ref. \cite{Doukourb1982},
and (b) DFT and two additional sets of manually tuned parameters. The exact field strengths are estimated based on experimental moment $M=9.5$ $\mu_B$.} 
\end{figure}

\begin{table}
\renewcommand*{\arraystretch}{1.4}

\begin{tabular}{ccccc}
\hline 
 $i$NN & distance (\(\text{\AA}\))  & $J_{i}^{\mathrm{DFT}}$ & $J_{i}^{\mathrm{set 1}}$ & $J_{i}^{\mathrm{set 2}}$\tabularnewline
\hline 
\hline 
1 & 4.186 & 1 & 1 & 1\tabularnewline
2 & 7.250 & 0.041 & 0.041 & 0.170\tabularnewline
3 & 8.372 & -0.190 & -0.100 & -0.019\tabularnewline
4 & 11.075 & -0.057 & -0.057 & -0.025\tabularnewline
\hline 
\end{tabular}


\caption{\label{tab:jex}First to fourth NN interactions and their distances d.
$J_{i}^{\mathrm{DFT}}$ ($i=1-4)$ are obtained by fitting into DFT
calculations. $J_{i}^{\mathrm{set 1}}$ and $J_{i}^{\mathrm{set 2}}$ are tuned manually around
the DFT parameters to explore the possible hidden steps. 
All values are divided by $J_{1}^{\mathrm{DFT}}=0.285$
}
\end{table}

To gain more insight, we look into a realistic system ErGa$_{2}$
by incorporating first-principles calculations. Figure \ref{fig:erga2}(a)
presents a comparison between the M-H curves taken from the prior
experimental data recorded at $1.5$ K \cite{Doukourb1982} and derived
using the parameters, $J_{i}^{\mathrm{DFT}}$, extracted from the DFT
calculations as listed in TABLE \ref{tab:jex}. At first glance,
these results exhibit outstanding agreement. Both featuring two distinct,
well-defined steps of $M=0.5$ and $1$ that correspond to phase \#3
and \#4 respectively that are induced at nearly identical applied
field strengths $H_{1}$ and $H_{2}$. 

However, upon closer inspection, two questions arise. Firstly, owing
to the strong easy-axis anisotropy on the Er sites, one would anticipate
that transitions occur as abrupt, vertical jumps once the magnetic
field reaches critical strength. Surprisingly, what we observe are
more gradual transitions. Secondly, the experimental data also reveal
narrow yet distinct hysteresis loops during both transitions.

Due to the frustrated nature of the system, as established in the
earlier discussions, there are potentially other states either degenerate
or very close in energy at the phase boundaries depending. It is reasonable
to suspect that these transitions are not necessarily direct, but
instead involving some intermediate states within very narrow windows
of the external magnetic field, giving rise to the seemingly more graduate
transitions and the formation of hysteresis loops. 

Indeed by observing the pattern in Fig. \ref{fig:j_rkky1}(b) and
exploring parameters near our DFT results, we managed to unveil two
additional steps, one corresponds to \#7 and the other to \#2 located
within a very narrow range of field around $H_{1}$, simply by making
a minor adjustment to $J_{3}$ from $-0.19$ to $-0.1$ as listed 
in TABLE \ref{tab:jex} and labeled $J_{i}^{\mathrm{set 1}}$.

With a slightly larger deviation from the $J_{i}^{\mathrm{DFT}}$
parameter set, we were also able to find another set of parameters
$J_{i}^{\mathrm{set 2}}$ that produce additional short intermediate state between
the \#3 to \#4 transition. Although the transition fields no longer
coincide as perfectly, the qualitative trend remains the same (i.e.
$H_{2}\sim3H_{1}$). Based on our analysis, we believe that in a more
accurate experiment, it is possible that the two additional steps
can occur.

To understand the possible origin of hysteresis, we discuss the issue
from two different aspects. If the intermediate states \#7 and \#2
exist as we predicted, then the direct transition from \#1 to \#7
or \#2 to \#3 are prohibited, as the transitions require complicated
multiple spin-flips all at once and one needs to grow a domain of
a new phase. This naturally implies hystereses.

On the other hand, geometrically, non-hysteretic transitions
are formally possible between \#1 and \#3, as well as between \#3 and \#4,
since these transitions can be achieved by sequentially flipping
the spin one by one. For these transitions, in order to establish the presence or absence of hystereses, one needs to check the critical field necessary to flip one spin, compare to the field needed to flip all spins in question.

For instance, for \#1 to \#3 transition, one can simply estimate the energy
cost of flipping one spin in the entire lattice under the field $H'_{1}$,
which has the analytic energy expression 
\begin{equation}
   E_{1+\mathrm{1flip}}-E_{1}=4J_{1}+4J_{2}-12J_{3}+8J_{4}-2H_{1}'
   \label{e1}
\end{equation}
and for the full transition at $H_{1}$ we have
\begin{equation}
E_{3}-E_{1}=4J_{1}+4J_{2}+8J_{4}-2H_{1}.
   \label{e2}
\end{equation}
Similarly, for the \#3 to \#4 transition, we have the following for the
energy cost of flipping one spin in \#3 under $H'_{2}$
\begin{equation}
H_{3+\mathrm{1flip}}-H_{3}=12J_{1}+12J_{2}-12J_{3}+24J_{4}-2H{}_{2}'
   \label{e3}
\end{equation}
and the energy for the full transition
\begin{equation}
H_{4}-H_{3}=12J_{1}+12J_{2}+24J_{4}-2H_{2}.
   \label{e4}
\end{equation}
Interestingly, the difference between Eqs. \ref{e1} and \ref{e2}, as well as between Eqs. \ref{e3} and \ref{e4}, depend only
on $J_{3}$. Using our DFT parameter $J_{3}^{\mathrm{DFT}}$,
we find that flipping one single spin in both cases requires a larger
field than triggering the full transitions (i.e. $H_{1}'>H_{1}$ and
$H_{2}'>H_{2}$), as a result, both transitions are predicted to be
hysteretic.

\section{Conclusions}

In conclusion, we have studied the effect of further-neighbors on metamagnetic transitions in classical Ising model on the triangular lattice. We determine the phase diagrams for two scenarios, through a comprehensive examination of energy comparisons among the most common possible magnetic states as well as the less intuitive ones proposed in the literature. The first scenario, where all the interactions are antiferromagnetic, is expected to host the richest phase diagrams and we discussed how the degeneracies are lifted with the presence of further-neighbors. The second one, motivated by our DFT analysis for ErGa$_{2}$, is introduced to mimic the Ruderman-Kittel-Kasuya-Yosida (RKKY) type of interaction where the sign of interactions vary with distance. Furthermore, we present a case study on a real-world system ErGa$_{2}$. By incorporating our first-principles-based exchange parameters into the I-3 model, we were able to accurately reproduce the experimentally observed transition steps. With the help of the analysis for the second scenario, we also predict the possible additional transition steps that could explain the discrepancy between the experimental data and the theory.




\begin{acknowledgments}
We are grateful to P. Nikolic and G. Schwertfeger for useful discussions.
\end{acknowledgments}




\bibliographystyle{apsrev4-1}
\bibliography{main}

\end{document}